# Palladium Diselenide Long-Wavelength Infrared Photodetector with High Sensitivity and Stability

Mingsheng Long[1, 4†], Yang Wang[1†], Peng Wang[1, 4†], Xiaohao Zhou[1, 4], Hui Xia[1, 4], Chen Luo[2], Shenyang Huang[3], Guowei Zhang[3], Hugen Yan[3], Zhiyong Fan[5], Xing Wu[2*], Xiaoshuang Chen[1, 4*], Wei Lu[1, 4], and Weida Hu[1, 4*]

1 State Key Laboratory of Infrared Physics, Shanghai Institute of Technical Physics, Chinese Academy of Sciences, 500 Yu Tian Road, Shanghai 200083, China.
2 Shanghai Key Laboratory of Multidimensional Information Processing, Department of Electronic Engineering, East China Normal University, 500 Dongchuan Road, Shanghai 200241, China.
3 Department of Physics, State Key Laboratory of Surface Physics and Key Laboratory of Micro and Nano Photonic Structures (Ministry of Education), Fudan University, 220 Han Dan Road, Shanghai 200433, China.
4 University of Chinese Academy of Sciences, 19 Yu Quan Road, Beijing 100049, China.
5 Department of Electronic and Computer Engineering, The Hong Kong University of Science and Technology, Clear Water Bay, Kowloon, Hong Kong, China SAR.

**ABSTRACT:**

A long-wavelength infrared (IR) photodetector based on two-dimensional materials working at room temperature would have wide applications in many aspects in remote sensing, thermal imaging, biomedical optics, and medical imaging. However, sub-bandgap light detection in graphene and black phosphorus has been a long-standing scientific challenge because of low photoresponsivity, instability in the air and high dark current. In this study, we report a highly sensitive, air-stable and operable long-wavelength infrared photodetector at room temperature based on $PdSe_2$ phototransistors and its heterostructure. A high photoresponsivity of ~42.1 $AW^{-1}$ (at 10.6 μm) was demonstrated, which is an order of magnitude higher than the current record of platinum diselenide. Moreover, the dark current and noise power density were suppressed effectively by fabricating a van der Waals heterostructure. This work fundamentally contributes to establishing long-wavelength infrared detection by PdSe$_2$ at the forefront of long-IR two-dimensional-materials-based photonics.

Scalable two-dimensional, long-wavelength infrared photodetectors operating at room temperature are highly desirable for upcoming remote sensing, thermal imaging, biomedical optics, medical imaging, and space communication applications. State-of-the-art long-wavelength infrared (LWIR) photodetectors based on narrow-bandgap semiconductors using HgCdTe alloy and III-V compound quantum structures suffer from several major challenges, such as the need for operation at liquid nitrogen temperatures, the complexity of sample synthesis and challenging device fabrication processes.[1] Commercial widely used LWIR photodetectors with 5-20 nm wavelength operating at room temperature based on VO$_x$ and α-Si possess many advantages such as compatibility with mass production, low price, and facile fabrication processes. However, their low sensitivity, short detection wavelength range and low response speed restrict their application.[2] Recently, the discovery of graphene, a two-dimensional layered material, has offered an opportunity to overcome some of these issues. In previous studies, LWIR photodetectors based on a graphene nanoribbon,[3] graphene quantum dot-like arrays[4] and a graphene heterostructure[5] have been demonstrated. Generally, the photoresponsivity has been low, approximately 7.5 μA W$^{-1}$ in the graphene nanoribbon, due to the limited light absorption of 2.3% in an atomic thin layer,[6] and a high dark current due to the gapless band structure. Although strategies such as surface plasma enhanced light absorption[7] and carrier multiplication[8-10] have been adopted to enhance the photoresponsivity of graphene photodetectors, the photoresponsivity is still relatively low at several tens of mA W$^{-1}$. A photoresponsivity of up to 0.4 AW$^{-1}$ at 10.6 μm was demonstrated by etching graphene to form quantum-dot-like arrays.[4] The resulting high responsivity was obtained at the cost of long response time. Notably, high-performance mid-IR detectors based on black phosphorus (b-P)[11, 12] were demonstrated due to its narrow bandgap of ~0.3 eV.[13, 14] Up to now, the operating spectral range of b-P photodetectors has been tuned to 7.7 μm based on a vertical electric field b-P device.[15] Notably, recently discovered black arsenic phosphorus (b-AsP), with the fraction of As increased to 83%, shows that the bandgap can be narrowed to ~0.15 eV.[16] Ultrabroadband photodetection based on a b-As$_{0.83}$P$_{0.17}$ phototransistor[17] covering the spectral range of ~8-14 μm (which extends to the second atmospheric transmission window) was demonstrated. The peak responsivity was as high as 17 AW$^{-1}$, and the cutoff wavelength reached 4.6 μm by using a b-AsP alloy-based device.[18] The current record of LWIR (~10 μm) photoresponsivity of ~4.5 AW$^{-1}$ was demonstrated based on platinum diselenide.[19] Broadband IR detection was also recently demonstrated for PtSe$_2$,[20] in addition to graphene and BP. However, black phosphorus is air sensitive.[21-23] The device fabrication process has to be carried out in a glove box filled with high-purity inert gas, and the device measurements must be conducted in a sealed environment or carried out in the vacuum. We summarize the performance of the LWIR photodetector in the supplementary material for those two-dimensional materials and conventional III-V and II-VI and HgCdTe materials in Table 1.

Layered materials with high infrared light absorption, high carrier mobility, and satisfactory stability have yet to be discovered. Theoretical calculation results predict that group X transition metal dichalcogenides (TMDs) (Ni, Pd, Pt) are promising narrow bandgap semiconductors with ~0-0.25 eV[24-28] in the bulk and with high room-temperature mobility[29, 30] greater than 1000 cm$^2$V$^{-1}$s$^{-1}$. Carrier mobility larger than 200 cm$^2$ V$^{-1}$ s$^{-1}$ and air-stable properties[24, 31] of the group X TMDs have been demonstrated in recent years. A high photoresponsivity of ~1560 AW$^{-1}$ in the visible range was demonstrated based on PtS$_2$ using h-BN as a substrate.[32] However, LWIR (8-14 μm) photodetection based on group X TMDs awaits further study.

Here, we report an experimentally synthesized PdSe$_2$ using a high-quality layered single crystal. An air-stable photodetector based on PdSe$_2$ FETs and PdSe$_2$-MoS$_2$ heterostructures operating at room temperature and at LWIR (up to 10.6 μm) were demonstrated. The photoresponsivity of the photodetector is as high as 42.1 AW$^{-1}$, which is an order of magnitude higher than the current record for PtSe$_2$ photodetectors. The specific detectivity $D^*$ is as high as $8.21 \times 10^9$ Jones under the illumination of a 10.6 μm wavelength infrared source in ambient air.

**RESULTS AND DISCUSSION**

The PdSe$_2$ layered material has been previously predicted to have excellent optoelectronics properties such as an extraordinarily high carrier mobility,[26, 30] large bandgap tenability from bulk to monolayer,[24] strong interlayer coupling and a narrow bandgap.[24, 25, 33] The crystalline structure of PdSe$_2$ is a pentagonal structure that is stable for only a few 2D materials.[34] Figure 1a shows a sketch of the PdSe$_2$ crystal structure in top view (top) and side view (bottom). The unit cell of bulk PdSe$_2$ is an orthorhombic structure with the space group Pbca (no. 61) and $D_{2h}$ point group symmetry.[26, 35, 36] Relative to well-studied layered TMDs, the best difference is that one palladium atom is coordinated with four selenium atoms, unlike the six coordinated transition metal atoms in typical 1T and 2H structures.

We calculated the electron band structure of bulk and few-layer PdSe$_2$ initially by *ab initio* calculation. Figure S1 presents the band structure of monolayer, bilayer, trilayer and bulk PdSe$_2$. For the bulk form, the valence band maximum (VBM) is situated at the high-symmetry $\Gamma$ (0, 0, 0) point, while the conduction band minimum (CBM) is situated between the $S$ (0.5, 0.5, 0) and $Y$ (0, 0.5, 0) points, exhibiting a 0.05 eV indirect bandgap. This result is consistent with previously reported results of ~0.03 eV.[25, 30] For the band structures of the monolayer and bilayer materials, the calculated results indicate indirect band gaps of ~1.23 eV and 0.85 eV, respectively, which is different from that of other widely studied TMDs exhibiting an indirect-to-direct band structure transition as the layer number transitions from bilayer to monolayer.[37] The multilayer PdSe$_2$ with a narrow bandgap less than 0.1 eV is a promising candidate for long-wavelength infrared photodetection.

PdSe$_2$ single crystals were obtained by a self-flux method, and the detail sees method. A high-purity Pd rod (4 mm diameter) (99.95%) and Se powder (99.999%) (200 mesh), all were purchased from Alfa Aesar. The $PdSe_2$ polycrystalline powder was prepared by solid-state reaction method in an evacuated quartz tube. The $PdSe_2$ single crystal was synthesized by a self-flux method using Se as a fluxing agent in a mass ratio of $PdSe_2$: Se =1: 4. A smooth surface of the $PdSe_2$ single crystal was obtained by cleaving the flake as shown in Fig. S2a. Raman spectroscopy was used to characterize the multilayered and bulk $PdSe_2$. The wavelength of the Raman exciting laser was 514 nm. As shown in Fig. 1b, four distinct Raman peaks were located at ~143, ~206, ~222 and ~256 $cm^{-1}$ corresponding to the $A_g^1$-$B_{1g}^1$, $A_g^2$, $B_{1g}^2$ and $A_g^3$ modes,[24, 25] respectively. The structure of the $PdSe_2$ phases was also confirmed by the X-ray diffraction (XRD) pattern as shown in Fig. 1c. The peaks located at 23.1°, 34.8°, 47.5° 50.2°, 65.0° and 74.1° can be indexed to the (002), (210), (300), (213), (400) and (006) planes, respectively. The peak intensity of (002) and (006) is extremely strong, indicating that $PdSe_2$ is a layered material along the *c* direction. The chemical composition of the $PdSe_2$ samples was confirmed by energy dispersive X-ray spectroscopy (EDXs) as shown in Fig. 1d. The high carbon and copper peaks (see Fig. S2b) originated from the carbon film and copper mesh of the transmission electron microscopy (TEM) sample holders. The atomic ratio Pd: Se is 33.05: 66.95, which is very close to 1: 2 as shown in the inset of Fig. 1d.

The crystalline structure of $PdSe_2$ was characterized by high-resolution transmission electron microscopy (HRTEM). The low-magnification TEM image is presented in Fig. S2c, where the scale bar is 100 nm. The HRTEM image of the (002) plane is shown in Fig. 1e and is consistent with the crystalline structure along the c-axis. The inset on the left corner of Fig. 1e is a high-resolution image obtained by the inverse fast Fourier transform. The selected area electron diffraction (SAED) patterns are very clear and further confirm the high quality of the single crystal as shown in Fig. 1f.

The electrical transport properties of few-layer $PdSe_2$ FETs were investigated. The device was fabricated by a conventional electron-beam lithography process.

Following standard electron-beam evaporation was used for fabricating of metal electrodes (5 nm Ti and 50 nm Au). Figure 2a presents the atomic force microscopy (AFM) image of a typical FET device, where the scale bar is 5 μm. The height profile is ~14 nm along the white dashed line. We chose PdSe$_2$ flakes with a thickness of ~5-20 nm for device fabrication because the highest mobility can be obtained at a thickness of ~10 nm[24, 25, 38, 39] and a relatively high light absorption can be achieved. Figure 2b and 2c plot the *I-V* curves and transfer curves of a typical FET device before and after annealing, respectively. An optical photograph of the measuring device is presented in the inset of Fig. 2b, where the scale bar is 5 μm. The annealing experiment was carried out at ~300°C with an argon flow of 200 standard cubic centimeters per minute (sccm) for 1 h in a tube furnace. According to the linear *I-V* curves, the contact between PdSe$_2$ and the metal electrode shows satisfactory ohmic contact. The annealing treatment can improve the mobility of the sample considerably.[25] The resistance at $V_g$ = 0 V increases from 0.34 MΩ to 0.81 MΩ after the annealing treatment. The transfer curves show mild hysteresis when the sweeping gate bias direction is changed. The dip points of the transfer curves are located at ~3.5 V and ~-1 V for the decreasing and increasing sweep of the gate bias, respectively. The dip points, similar to the 'Dirac point' of graphene, are quite close to 0 V, which indicates that the samples are intrinsic without doping. The carrier mobility can be calculated by $\mu = \frac{L}{WC_gV_{ds}} \times \frac{dI_{ds}}{dV_g}$, where *L* and *W* denote the channel length and width, respectively. $C_g$ = 11.5 nF cm$^{-2}$ is the capacitance per unit area of the 300 nm SiO$_2$. The electron and hole maximum mobilities are ~59.8 and 16.1 cm$^2$V$^{-1}$s$^{-1}$ before annealing, respectively, with an on/off ratio of ~10$^2$. After the annealing experiment, the electron and hole mobilities increase to ~138.9 and 57.0 cm$^2$V$^{-1}$s$^{-1}$, respectively, which is slightly smaller than the previously reported[25] $\mu_{e\ (max)}$~216 cm$^2$V$^{-1}$s$^{-1}$. In addition, the on/off ratio is increased to 10$^3$, and the hysteresis decreases. Furthermore, the dip points of the transfer curves are close to 0 V after the annealing treatment. This treatment can enhance the sample quality by driving off the surface absorbed states[25] and repairing defects that originate from sample exfoliation and device fabrication. A p-type transport behavior was also achieved according to the transfer characteristic curves (see Fig. S3a). Moreover, we also used Pd/Au as a contact to obtain p-doping $PdSe_2$ to study the performance of $PdSe_2$ phototransistor.[40] Pd/Au as contact electrodes, a higher Schottky barrier was formed which can depress the dark current. The large contact resistance also decreased the photoresponse due to the inefficiency photocarrier collection.

To investigate the spectral photoresponse of $PdSe_2$ in the long-wavelength infrared spectral range, we measured the optical absorption spectrum (see Fig. S3b) using a typical multilayer $PdSe_2$ sample. As the wavenumber decreases, the absorption spectrum decreases linearly to ~650 $cm^{-1}$ (corresponding to ~15.4 μm) and is marked by the crossing of the two dashed cyan lines (see Fig. S3b) for two thicknesses of $PdSe_2$ sheets of 30 nm (Fig. S3b inset) and 150 nm (See Fig. S3c). Note that the absorption edge of multilayer $PdSe_2$ is located at approximately 650 $cm^{-1}$, which indicates that the samples can absorb light wavelengths longer than 15 μm. To reveal the photoresponse in the long-wave IR, we fabricated a field effect transistor (FET) using the multilayer $PdSe_2$ flakes. Fig. 3a presents the linear output curves of a typical $PdSe_2$ phototransistor with and without illumination (10.6 μm). The spot size of the laser is ~3 mm in diameter, which is much larger than our device's channel length of ~10 μm. Thus, the devices were fully illuminated. The photocurrent under laser illumination was detectably larger than that under dark conditions. The optical image of the measured $PdSe_2$ phototransistor is presented in the inset of Fig. 3a. To further evaluate the photoresponse in the LWIR range, the time-resolved photoresponse at 1 V bias under switched illumination was measured. With illumination, the photocurrent increases sharply and saturates. Three cycles were measured as shown in the inset of Fig. 3b. The highly repeatable photocurrent generation reveals that the photoresponse of our devices is reversible and stable. We extracted the photoresponsivity ($R$), one of the crucial parameters of a photodetector, defined as $R = I_P / P_I$, where $I_P$ is the photocurrent and $P_I$ is the incident light power. Fig. 3b presents the extracted light-power-dependent photoresponsivity and photoconductivity gain ($G$) at 1 V bias. Notably, a photoresponsivity of up to ~42.1 AW$^{-1}$ is obtained at $V_{ds}$ = 1 V bias under laser illumination of 10.6 μm. The photoresponsivity decreases from 42.1 to 13.8 AW$^{-1}$ as the light power is increased from 1.42 nW to 56.7 nW. To evaluate the multiplication of photogenerated carriers, we calculated $G$, which can be expressed as $G = (h\nu/q)(R/\eta)$, where $h\nu$ is the photon energy and $\eta$ is the photon absorption efficiency. The photogating effect plays a crucial role in high photoconductive gain due to the long lifetime of combination induced by the trap state and short carrier transit time.[11] The calculated $G$ is as high as ~49 and decreases to 16 as the illumination power increases if $\eta$ = 10% is assumed (see Fig. S3b). To further examine the broadband response of our photodetectors, we measured the photoresponse from 450 nm to 10.6 μm at a 1 V bias. The extracted wavelength-dependent photoresponsivity and photoconductive gain are plotted in Fig. 3c. In the visible range, the photoresponsivity decreases sharply from ~249.1 AW$^{-1}$ to ~17.7 AW$^{-1}$ as the incident spectrum of light increases from 450 nm to 940 nm. At the same time, the response time also decreases quickly (see Fig. S4), which is attributed to the long lifetime of the high-energy trap states. As the wavelength increases from near IR to LWIR up to 10.6 μm, the responsivity increases slowly from ~17.7 AW$^{-1}$ to ~37.7 AW$^{-1}$ at the shortwave infrared range (1-3 μm) and becomes stable at ~45 AW$^{-1}$ with little fluctuation from the mid-wave infrared range (3-5 μm) to the LWIR range (8-12 μm). Correspondingly, the gain decreases sharply from 6861 to 344 as the wavelength increases from 450 nm to 637 nm. Then, the gain decreases slowly to 49 as the wavelength increases to 10.6 μm. We also measured the time-resolved photoresponse at various wavelengths with a source-drain voltage of 1 V. From 450 nm to 940 nm, the photocurrents range from several tens of μA to several μA (see Fig. S4). One of the most important figures of merit for photodetectors is the photoresponse speed. We next measured the response time under 10.6 μm laser illumination. The rise/decay time is defined as the time required to transition from 10/90% to 90/10% of the stable photocurrent during the illumination on/off cycle. A rise time of $\tau_{rise}$ = 74.5 ms and a decay time of $\tau_{decay}$ = 93.1 ms are obtained (see Fig.

S5a). The response time under an illumination of 637 nm laser is much faster. A rise time of $\tau_{rise}$ = 51.3 μs and a decay time of $\tau_{decay}$ = 53.7 μs, which are more than two orders of magnitude faster than that at LWIR range, are achieved as shown in Fig. S5b. The output curve in an illuminated condition is much higher than that in a dark condition (see Fig. S5c), which indicates that a large photocurrent is generated when the illumination is present. We also measured the power-dependent photoresponsivity from 2.7 μm to 10.6 μm (see Fig. S5d). Notably, the illumination power dependence of photoresponsivity at 2.7, 3.0 μm, 4.012 μm, and 10.6 μm exhibits a similar trend, and the values are similar, which indicates that the mechanism of dominant photocurrent generation in the range of 2.7 μm to 10.6 μm is the same. The photoresponsivity at four different illumination levels ranges from ~13.8 to ~39.7 AW$^{-1}$ as the incident illumination power ranges from ~1.4 nW to ~70.8 nW. To explore the mechanism of the photoresponse in the long-wavelength infrared (10.6 μm), we measured the transfer curves under dark conditions and various incident light powers as shown in Fig. S5e. As the incident light power increases, the troughs of the transfer curves shift horizontally to the right. During illumination, the trap states at the defects or at the interface trap one type of photocarrier for an extended time. In our PdSe$_2$ phototransistor, the hole is trapped. The electric field effect of the trapped holes can induce the movement of more electrons in the channel, which can reduce the resistance and allow additional current flow. The illumination-induced horizontal shift of the transfer curve is a typical characteristic of the photogating effect.[41, 42]

To examine the sensitivity of PdSe$_2$ phototransistors, we measured the current-noise density spectra at $V_{ds}$ = 1 V. The noise spectra of the PdSe$_2$ phototransistor at 1 V bias are shown in Fig. S5f. At the low-frequency point, 1/$f$ noise dominates the noise current contribution. The low-frequency flicker noise originates from the fluctuation of carriers being trapped and de-trapped by defects and disorder,[43] which exist widely in 2D materials.[44, 45] We then calculated another important figure of merit, the noise equivalent power (NEP), which is related to the sensitivity of a photodetector. The NEP is defined as $i_n$ / R, where $i_n$ is the measured noise current. The NEP of the PdSe$_2$ phototransistor at the full range from 0.45 μm to 10.6 μm is lower than 0.28 pW Hz$^{-1/2}$ as shown in Fig. 3d. The specific detectivity, $D^*$, which is used to evaluate the sensitivity of a photodetector, is the minimum signal power that a photodetector can distinguish from noise. The wavelength-dependent specific detectivities $D^*$ are plotted in Fig. 3d. Notably, at 10.6 μm, $D^*$ is as high as $1.10 \times 10^9$ Jones. These values are well beyond the current records of 2D-based LWIR photodetectors at room temperature[46] and even better than that of RT-operated HgCdTe, PbSe, and InSb photoelectromagnetic (PEM) detectors and commercial thermistor bolometers.

The high dark noise current density remains a major limitation for narrow-bandgap-semiconductor and semimetal-based photodetectors. Note that the PdSe$_2$ FET device shows weak p-type conduction (see Fig. S3a). The p-n junctions could be fabricated by stacking weak p-type PdSe$_2$ with other n-type 2D layered materials to suppress the dark current. Here, the widely used strategy to suppress the dark current noise by fabricating a 2D van der Waals (vdW) heterostructure was adopted. The built-in electrical field at the junction can significantly reduce the dark noise current. The PdSe$_2$ material shows a weak p-type semiconductor property. We chose n-type MoS$_2$ to deposit on PdSe$_2$ to form a vdW heterostructure junction. A schematic image of the PdSe$_2$-MoS$_2$ heterostructure infrared photodetector is shown in the top panel of Fig. 4a. The bottom panel of Fig. 4a is the optical image of the heterostructure device. The scale bar is 5 μm. Under dark conditions, the output curve shows a satisfactory rectification effect. The rectification ratio of a typical heterostructure device at 2 V bias is as high as 10$^2$, which indicates that a built-in electrical field exists at the interface. Then, we explored the broadband photovoltaic response of a PdSe$_2$-MoS$_2$ heterostructure. The photoresponse of the device under a 10.6 μm illumination is presented in Fig. 4b. The photocurrent under illumination is substantially higher than that under dark conditions. The time-resolved photoresponse of a typical PdSe$_2$-MoS$_2$ heterostructure device at $V_{ds}$ = 1 V under 10.6 μm is plotted in the inset of Fig. 4b; a photocurrent of more than 80 nA is obtained. We also studied the photovoltaic response of the heterostructure. The time-resolved photovoltaic response under 940 nm laser illumination was measured, revealing a 47.4 nA photocurrent (see Fig. S6a). The *I-V* curves of a typical PdSe$_2$ phototransistor with and without illumination on of a 940 nm laser (see Fig. S6b). The extracted photovoltaic *R* decreases from ~185.6 to ~32.3 mA W$^{-1}$, and the calculated external quantum efficiency (EQE) decreases from 24.5% to 4.3% as the illumination power increases from 4.5 nW to 1.46 μW (see Fig. S6c). This heterostructure device shows a satisfactory photovoltaic response in the near-infrared wavelength range. The power dependence photoresponsivity for visible (637 nm) and near-infrared (940 nm) light for the PdSe$_2$-MoS$_2$ heterostructure device at $V_{ds}$ = 1 V is presented in Fig. S6d. The *R* is as high as 11.15 AW$^{-1}$ at 637 and 4.24 AW$^{-1}$ at 940 nm. Importantly, a broadband photoresponse from 450 nm to 10.6 μm is observed for the vdW heterostructure devices. Then, we extracted the photoresponsivity over the visible to LWIR wavelength range; the wavelength-dependent responsivities are presented in the left y-axis of Fig. 4c. The responsivity decreases sharply from ~22.86 AW$^{-1}$ to ~4.24 AW$^{-1}$ as the illumination wavelength increases from 450 nm to 940 nm. The responsivity then becomes stable at ~4 AW$^{-1}$ and shows only small fluctuations as the illumination wavelength increases to 10.6 μm. A photoresponsivity peak located at ~4 μm is observed. The responsivity peak at ~4 μm was also found for another heterostructure device (see Fig. S7a). We then measured the illumination-power-dependent photoresponsivity of the vdW device at various wavelengths in the IR (see Fig. S7b). Notably, the photoresponsivity at 4.012 μm is higher than that of the other three wavelengths ($\lambda$ = 2.7 μm, 3.1 μm and 10.6 μm). The photoresponsivity is as high as ~28.83 AW$^{-1}$ at 1 V bias under 4.012 μm laser illumination. The high photoresponsivity peak at ~4 μm can be attributed to the much higher photoconductive gain at weak light condition[11] and comparatively fairly high light absorption around ~4.5 μm (See Fig. S3d). The band alignment of MoS$_2$-PdSe$_2$ heterostructure is type I, and the VBM and CBM of multilayer MoS$_2$ are ~-5.84 eV and ~-4.25 eV, respectively.[47] The work function of PdSe$_2$ is ~5.4 eV[48], and the bandgap of multilayer $PdSe_2$ is ~0.1 eV (see Fig. S7c). Interlayer excitons[49, 50] are generated as the photons of $hv_3$~0.3 eV (corresponding to ~4.1 μm) and $hv_4$~1.2 eV (corresponding to ~1.0 μm) are incident on the heterostructure, enhancing the light absorption at ~4.1 μm and ~1.0 μm illumination, respectively, and enabling a high photoresponse. The interlayer excitons may be the origin of the photoresponsivity peak at ~4.012 μm and the sharp decrease in photoresponsivity as the wavelength increases to an almost stable value from the ~1.0 μm spectrum range. Notably, the photoresponse speed for $PdSe_2$-$MoS_2$ heterostructures is also very fast under 637 nm illumination with a rise time of ~65.3 μs and a decay time of ~62.4 μs as shown in Fig. S7d. The fast photoresponse can be attributed to fast charge transfer at the interface of the heterostructure.

To examine the sensitivity of the $PdSe_2$-$MoS_2$ heterostructure photodetectors, we measured the current-noise density spectra at $V_{ds}$ = 1 V. The current-induced noise level of the heterostructure is significantly depressed relative to that of the $PdSe_2$ FET, as expected. At the low-frequency point, $1/f$ noise dominates the noise current contribution. The low-frequency flicker noise originates from the fluctuation of carriers being trapped and de-trapped by defects and disorder,[43] which exist widely in 2D materials.[44, 45] Notably, as the frequency increases beyond 1000 Hz, the noise current of the heterostructure quickly decreases to the Johnson noise level. The Johnson noise is a white background noise, which can be expressed as $<i_n^2>$ = 4 $k_B$ $T$ $\Delta f$ / $R_0$, where $k_B$ is the Boltzmann constant, $T$ is the temperature, $\Delta f$ is the bandwidth and $R_0$ is the device resistance. We calculated $<i_n^2>$ = 3.2×10$^{-26}$ A$^2$ at $\Delta f$ = 1 Hz, $R_0$ = 0.52 MΩ and $T$ = 300 K, which is consistent with the experimental result, while for the $PdSe_2$ FET devices, the current-noise is three orders of magnitude higher than the Johnson noise level. This result indicates that the built-in electric field at the junction can effectively depress the noise level and is highly desired. The NEP of the heterostructure at the full range from 0.45 μm to 10.6 μm is lower than 0.13 pW Hz$^{-1/2}$ as presented in the right y-axis of Fig. 4c. The wavelength-dependent specific detectivities $D^*$ of the $PdSe_2$-$MoS_2$ heterostructure device and other traditional LWIR detectors are plotted in Fig. 4d. Over the full range, $D^*$ is greater than $6.88\times10^9$ Jones at room temperature. Notably, at 10.6 μm, $D^*$ is as high as $8.21\times10^9$ Jones. The peak of $D^*$ for the PdSe$_2$-MoS$_2$ heterostructure is located at 4.012 μm and reaches $6.09 \times 10^{10}$ Jones. These values are well beyond the current records for room-temperature-operated PtSe$_2$ 2D-based LWIR photodetectors (~$7 \times 10^8$ Jones)[19] and graphene thermopiles (~$8 \times 10^8$ Jones)[46] and even better than that of uncooled HgCdTe (295 K, peak ~$4 \times 10^8$ Jones)[51] InSb, PbSe and commercial thermistor bolometers. At the MWIR range, the detectivity is on par with the b-AsP-MoS$_2$ heterostructure.[17] Moreover, the PdSe$_2$ sample is very stable in ambient air. Fig. S8a and Fig. S8b shows an optical image of the PdSe$_2$ phototransistor; the device had been exposed in air for more than three months. There is hardly any degradation of the sample (as verified by the optical images) or a decrease in the photocurrent (see Fig. S8c and S8d) at the same incident light power. For the PdSe$_2$-MoS$_2$ heterostructure device, a sample was stored in a dry box filled with air for nearly one year. The device and photoresponse at 10.6 μm appear the same as a fresh device (see Fig. S8e-S8h). All measurements were carried out in ambient air. We also measured the Raman spectrum of PdSe$_2$ and the Raman and PL spectra of MoS$_2$ using a typical PdSe$_2$-MoS$_2$ heterostructure device that was exposed in are more 6 months (see Fig. S9a). The high quality of the Raman spectrum (see Fig. S9b) further confirms the stability of the PdSe$_2$ sample. The Raman spectra of the MoS$_2$ and PL devices indicate that the MoS$_2$ used in the device is a multilayer MoS$_2$ sheet (see Fig. S9c and S9d). We also use Pd/Au (5 nm/ 50 nm) and Cr/Au (5 nm/ 50 nm) as a contact. Good Ohmic contact is only obtained by using Ti/Au (5 nm/ 50 nm) as contact electrodes (see Fig. S10).

**CONCLUSION**

In summary, high-quality, narrow-bandgap and air-stable single-crystal PdSe$_2$ was obtained by the self-flux method. A photoresponsivity of up to 42.1 AW$^{-1}$ for PdSe$_2$ FET devices and a specific detectivity up to $8.21 \times 10^9$ Jones for PdSe$_2$-MoS$_2$ heterostructure devices was demonstrated at room temperature for LWIR 10.6 μm illumination. Relative to other infrared materials, such as graphene[3, 4] or b-AsP[17], $PdSe_2$ exhibits the significant advantages of high sensitivity, fast speed and stability in ambient air. The dark current and the current-noise density were sharply attenuated by forming vdW heterostructures. Further efforts may include growing large-scale and high-quality $PdSe_2$ crystalline thin films and developing a scalable fabrication technique for LWIR room-temperature imaging. Our results not only exemplify an ideal case for the challenging LWIR spectral range photodetector but also for LWIR technologies, such as LWIR room-temperature imaging.

**METHODS**

**Materials synthesis**

The $PdSe_2$ single crystal was obtained by the self-flux method. A high-purity Pd rod (4 mm diameter) (99.95%) and Se powder (99.999%) (200 mesh) were purchased from Alfa Aesar. A mixture of Pd and Se in an atomic ratio of Pd: Se = 1: 2 was sealed in an evacuated quartz tube at $10^{-3}$ Pa to grow the $PdSe_2$ poly-crystal powder. The sealed quartz tubes were placed in a tube furnace that was slowly heated to 800 °C and then held at that temperature for 5 h to complete the full reaction. Subsequently, the furnace was heated to 1050 °C within 2 h, and this temperature was held for 20 h before the furnace was switched off. The obtained $PdSe_2$ powder was mixed with Se power in a mass ratio of $PdSe_2$: Se =1: 4. The mixed powder was then resealed in an evacuated quartz tube, and the sample was placed in a box furnace that was slowly heated to 850 °C, held at that temperature for 70 h, and then slowly cooled down to 450 °C at a rate of 2 °C $h^{-1}$. Finally, the furnace was switched off, and the sample was allowed to cool to room temperature.

**Device fabrication and measurements**

Multilayer $PdSe_2$ samples were obtained by using a standard mechanical exfoliation method. Single-crystal $PdSe_2$ was exfoliated on $SiO_2$/Si substrates. The thickness of the $PdSe_2$ flakes was measured using an atomic force microscope (Bruker Multimode 8). The multilayer $MoS_2$ was peeled from commercially available crystal $MoS_2$ samples (SPI supplied) on polydimethylsiloxane (PDMS). $PdSe_2/MoS_2$ heterostructures were fabricated using a 'PDMS transfer' technique in ambient air. The metal electrodes (5 nm Ti/ 50 nm Au) were fabricated using an electron-beam lithography process followed by an electron-beam evaporation process.

The electrical transport behavior and photoresponsivity were characterized in ambient air. A highly sensitive Keithley 2636B dual channel digital source meter was used for applying the bias and gate voltages. A commercial $CO_2$ laser source ($\lambda$ = 10.6 μm) was used as a long-wavelength infrared light source during the photoresponse measurements. The spot size across the spectrum ranging from 2.7 μm to 10.6 μm was ~3 mm in diameter. The lasers were focused by a 20× objective lens over the visible to short wavelength infrared range (450 nm to 940 nm). Noise current density spectra at various bias voltages were measured in a thoroughly shielded box in ambient air. The data were acquired using a spectrum analyzer (SR770) with a 100 kHz measuring bandwidth. HRTEM analysis was carried out on a JEM2100F with an acceleration voltage of 200 kV. To avoid damage to the $PdSe_2$ samples, the e-beam was carefully defocused.

**DFT Calculations**

The $PdSe_2$ band structure calculations were performed using the density-functional theory (DFT) with the Vienna *ab initio* simulation package (VASP).[52, 53] The projector augmented wave method (PAW) was used to describe the electron-ion interaction. The kinetic energy cutoff for the plane waves was set to 400 eV with an energy precision of $10^{-5}$ eV. The electron exchange-correlation function was addressed using a generalized gradient approximation (GGA) in the form proposed by Perdew, Burke, and Ernzerhof (PBE).[54] Van der Waals (vdW) interactions between the $PdSe_2$ layers were considered using the vdW density functional method optPBE-vdW.[55] The GGA+U approach was employed in the calculations of the structural and electronic properties. Values for the *U* and *J* parameters are chosen as $U$ = 3.94 eV and $J$ = 0.59 eV, respectively.[56] Both atomic positions and lattice vectors were fully optimized using the conjugate gradient algorithm until the maximum atomic forces were less than 0.001 eV/Å. A 12×12×10 Monkhorst-Pack k-point mesh was used for the Brillouin zone sampling.

## ASSOCIATED CONTENT

**Supporting Information**

Supplementary material including first-principles calculated band structure of $PdSe_2$, EDX results of $PdSe_2$, light absorption spectrum, Raman spectra of $PdSe_2$ and $MoS_2$ sheets, PL spectrum of $MoS_2$, the noise current spectra and photoresponse of $PdSe_2$ phototransistors and $PdSe_2$-$MoS_2$ heterostructures.

The authors declare no competing financial interests.

## AUTHORS INFORMATION

**Corresponding Authors**

E-mail: wdhu@mail.sitp.ac.cn

E-mail: xschen@mail.sitp.ac.cn

E-mail: xwu@ee.ecnu.edu.cn)

**ORCID**

Mingsheng Long: 0000-0002-1646-7153

Xin Wu: 0000-0002-9207-6744

Weida Hu: 0000-0001-5278-8969

**Authors contributions:**

M. L. and W. H conceived the project and designed the experiments. M. L., Y. W. and H. X. performed device fabrication and characterization. C. L and X. W performed the TEM measurements. M. L., P. W. and W. H. performed data analysis. X. Z. performed the *ab initio* calculations. M. L. and W. H. cowrote the paper, and all authors contributed to the discussion and preparation of the manuscript. We thank Dr. Xiaowei Liu of the Nanjing University for experimental help, Prof. Xiaomu Wang and Prof. Feng Miao of the Nanjing University for useful discussions, and James Torley of the University of Colorado at Colorado Springs for critical reading of the manuscript.

†M. L., Y. W. and P. W. contributed equally to this work.

ACKNOWLEDGMENT

This work was supported in part by the National Natural Science Foundation of China (grant nos. 61725505, 61835012, 11734016, 61521005, and 61674157), Fund of Shanghai Natural Science Foundation (grant no. 18ZR1445800), Key Research Project of Frontier Science of Chinese Academy of Sciences (grant no. QYZDB-SSW-JSC031), Fund of SITP Innovation Foundation (cx-190) and CAS Interdisciplinary Innovation Team.

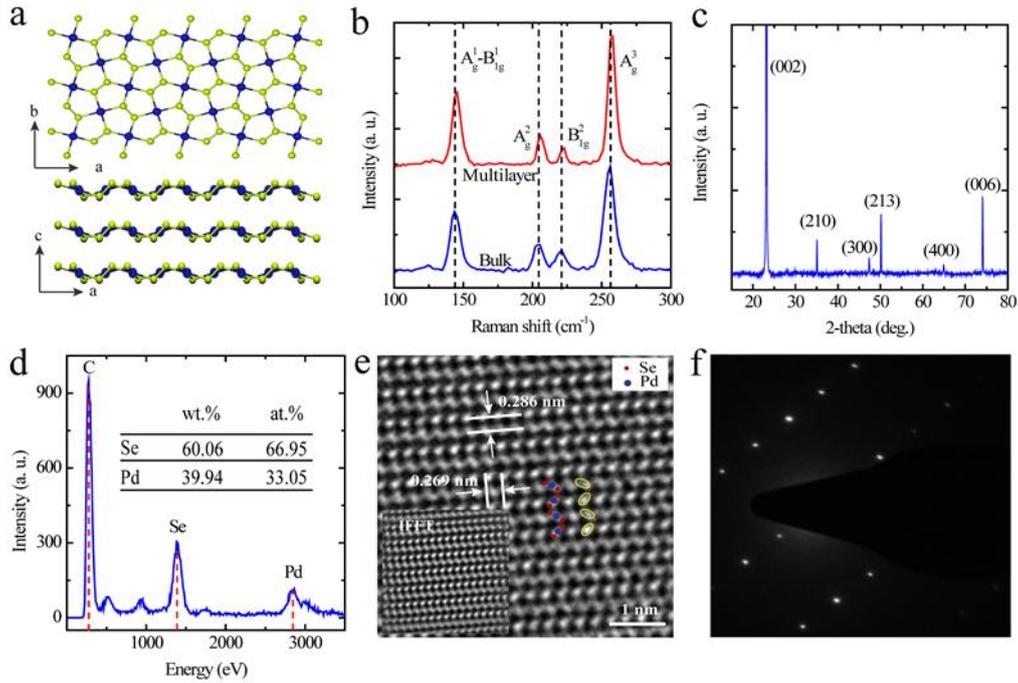

**Figure 1. PdSe$_2$ single crystal structure characterization and band structure calculation.** (a) Top panel: top view of the crystal structure of monolayer PdSe$_2$ Sheet. Bottom panel: side view of the crystal structure of multilayer PdSe$_2$ flake. (b) Raman spectra of bulk and multilayer PdSe$_2$. (c) X-ray spectra of a PdSe$_2$ single crystal flake. (d) Energy Dispersive X-ray Spectroscopy (EDX) of PdSe$_2$ flake. (e) High resolution transmission electron microscopy (TEM) image of the PdSe$_2$ (002) planes (f) Selected-area electron diffraction (SAED) pattern of the PdSe$_2$.

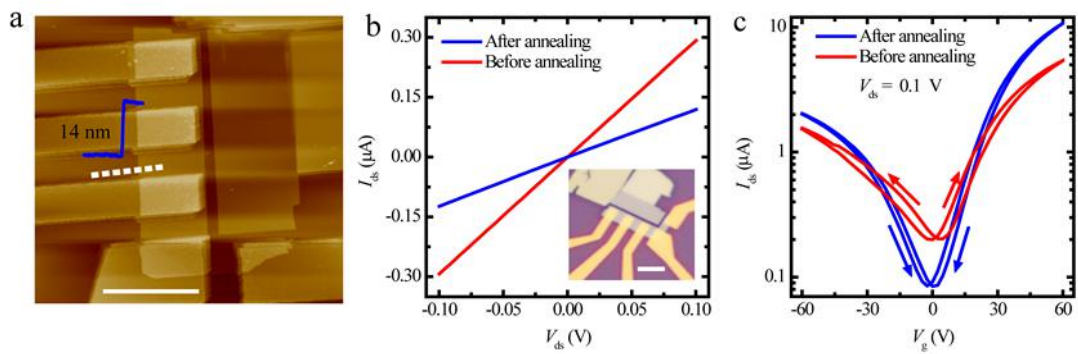

**Figure 2. Atomic force microscopic image and electric transport characterization of PdSe$_2$ a typical FET device.** (a) Atomic force microscopic image of the PdSe$_2$ FET device. The height profile is along the white dash line, scale bar 5 μm. (b) Output curves of typical FET device before and after annealing. Inset: optical image of the measured device, scale bar 5 μm.   (c) Transfer curves of the device before and after

annealing.

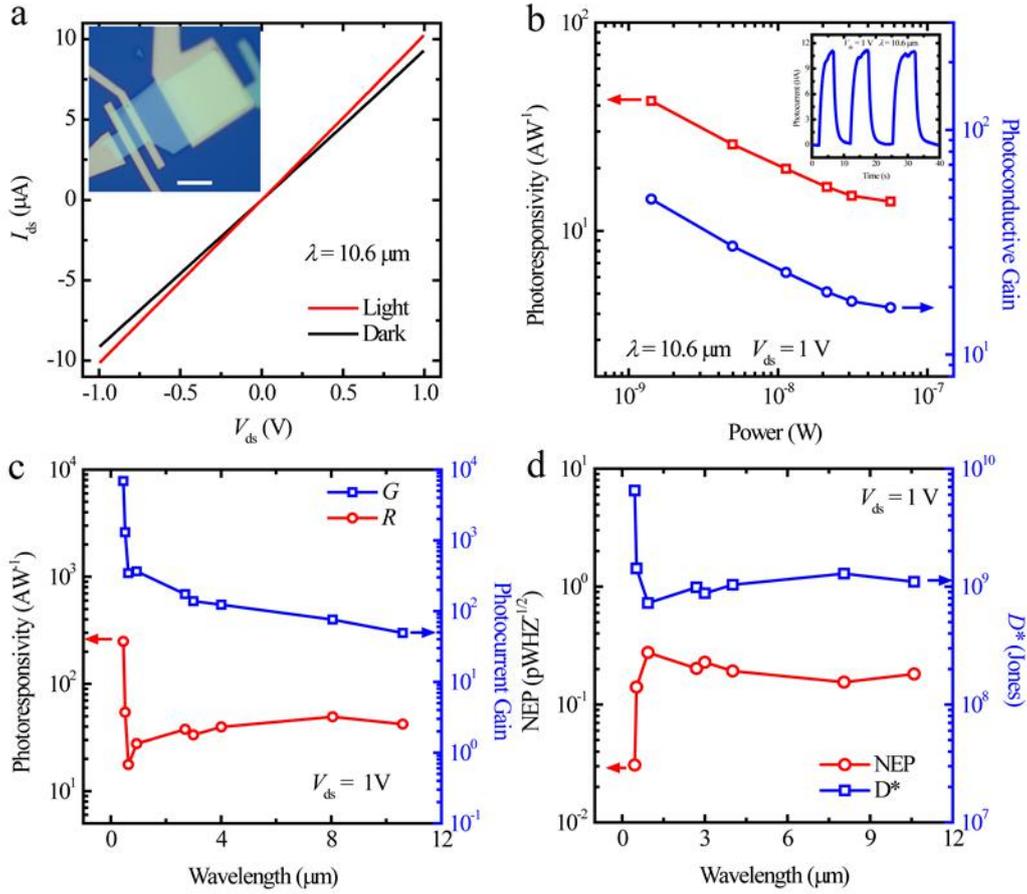

**Figure 3. LWIR photoresponse of a typical PdSe₂ phototransistor.** (a) Output curves of a typical PdSe$_2$ phototransistor with and without light illumination. The incident light power was 23.6 nW. Inset: optical image of the measured device, scale bar 5 μm. (b) The extracted power dependence photoresponsivity $R$ (left) and gain $G$ (right) at $V_{ds}$ = 1V. Inset: The time-resolved photoresponse under a 10.6 μm wavelength illumination at 1 V bias. The illumination power was fixed at 56.7 nW. (c) The extracted photoresponsivity $R$ (left) and $G$ (right) as a function light wavelength at 1 V bias in ambient air. (d) Wavelength-dependent noise equivalent power (red open circle) and specific detectivity $D^*$ (blue open square) of PdSe$_2$ phototransistor at $V_{ds}$ = 1 V in ambient air.

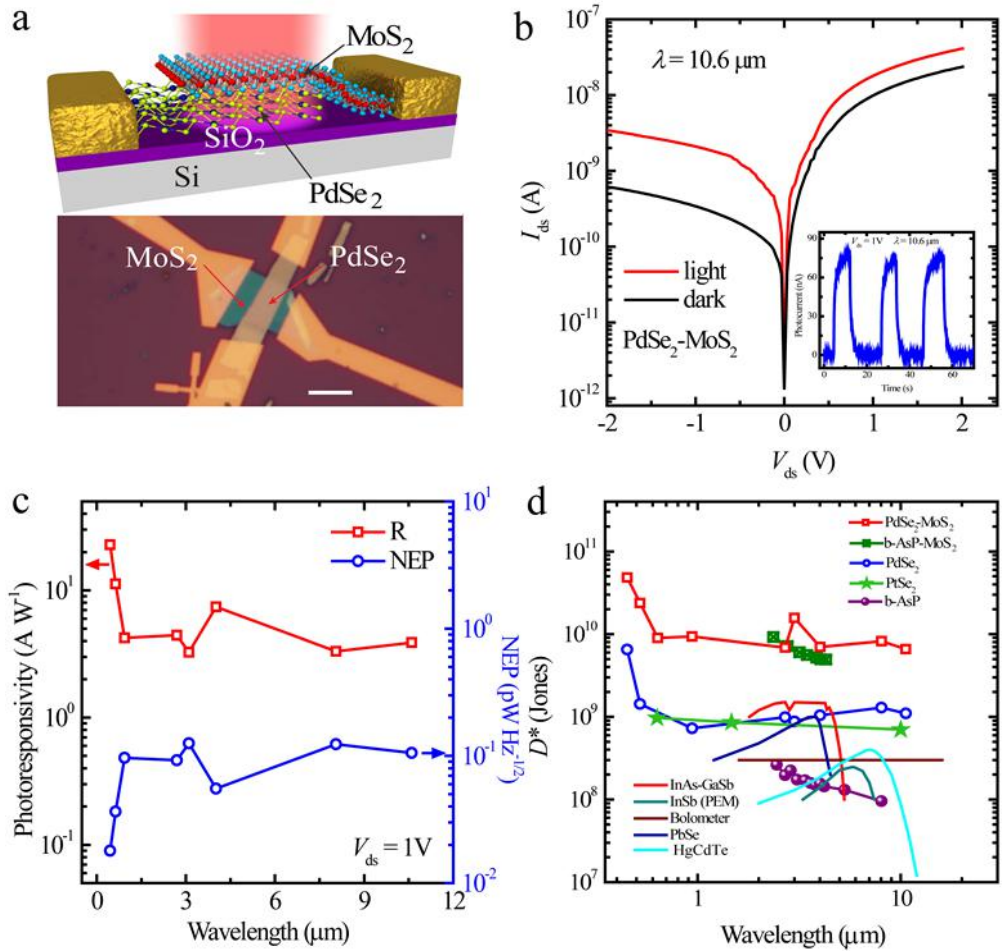

**Figure 4. High sensitivity and broadband photoresponse of PdSe$_2$-MoS$_2$ heterostructure device.** (a) Top panel: Schematic image of the PdSe$_2$-MoS$_2$ infrared photodetector. Bottom panel: optical photograph of the PdSe$_2$-MoS$_2$ device, scale bar 5 μm. (b) Semi-logarithmic plot of $I_{ds}$-$V_{ds}$ characteristic curves with and without the light on. The light power was fixed at 435.9 nW under a 10.6 μm laser. Inset: The time-resolved photoresponse of PdSe$_2$-MoS$_2$ photodetector under a 10.6 μm wavelength illumination at 1 V bias. (c) The extracted wavelength dependent photoresponsivity $R$ and noise equivalent power (blue open circle) of the PdSe$_2$-MoS$_2$ photodetector at $V_{ds}$ = 1 V in ambient air. (d) Room temperature specific detectivity $D^*$ as a function of wavelength for various 2D materials and conventional infrared materials.

TOC

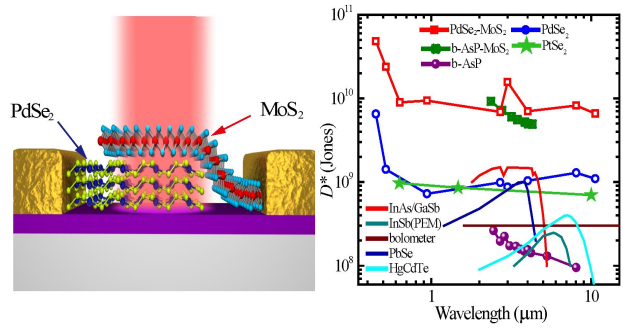